\documentclass[reprint,amsmath,amssymb,aps,prl,twocolumn,showpacs,superscriptaddress, floatfix]{revtex4-1}
\usepackage{amssymb}
\usepackage{graphicx}
\usepackage{dcolumn}
\usepackage{bm}
\usepackage{pslatex}
\usepackage{xcolor}

\begin{document}

%\preprint{}

\title{Unconventional Correlation between Quantum Hall Transport Quantization and Bulk State Filling in Gated Graphene Devices}


\author{Yong-Tao Cui}
\affiliation{Geballe Laboratory for Advanced Materials (GLAM), Stanford University, Stanford, CA 94305, USA}

\author{Bo Wen}
\affiliation{Department of Physics, Columbia University, New York, NY 10027, USA}

\author{Eric Y. Ma}
\affiliation{Geballe Laboratory for Advanced Materials (GLAM), Stanford University, Stanford, CA 94305, USA}

\author{Georgi Diankov}
\affiliation{Geballe Laboratory for Advanced Materials (GLAM), Stanford University, Stanford, CA 94305, USA}

\author{Zheng Han}
\affiliation{Department of Physics, Columbia University, New York, NY 10027, USA}

\author{Francois Amet}
\affiliation{Department of Physics and Astronomy, Appalachian State University, Boone, NC 28607, USA}

\author{Takashi Taniguchi}
\affiliation{National Institute for Materials Science, 1-1 Namiki, Tsukuba 305-0044, Japan}

\author{Kenji Watanabe}
\affiliation{National Institute for Materials Science, 1-1 Namiki, Tsukuba 305-0044, Japan}

\author{David Goldhaber-Gordon}
\affiliation{Geballe Laboratory for Advanced Materials (GLAM), Stanford University, Stanford, CA 94305, USA}

\author{Cory R. Dean}
\affiliation{Department of Physics, Columbia University, New York, NY 10027, USA}

\author{Zhi-Xun Shen}
\affiliation{Geballe Laboratory for Advanced Materials (GLAM), Stanford University, Stanford, CA 94305, USA}


\date{\today}

\begin{abstract}
We report simultaneous transport and scanning Microwave Impedance Microscopy to examine the correlation between transport quantization and filling of the bulk Landau levels in the quantum Hall regime in gated graphene devices. Surprisingly, comparison of these measurements reveals that quantized transport typically occurs below complete filling of bulk Landau levels, when the bulk is still conductive. This result points to a revised understanding of transport quantization when carriers are accumulated by gating. We discuss the implications on transport study of the quantum Hall effect in graphene and related topological states in other two dimensional electron systems.
\end{abstract}

%\keywords{}

\maketitle
The quantum Hall (QH) effect~\cite{Klitzing1980} has proven to be a powerful tool for understanding the physics of two dimensional electron systems (2DESs). The occurrence of transport quantization has been associated with the topological order of the Landau levels (LLs) in the bulk of a 2DES. When carriers completely fill a bulk LL, the Fermi level lies in the energy gap above the LL thus the bulk becomes highly incompressible~\cite{Chklovskii1992}. The transverse Hall conductance is determined by the total Chern number of the occupied LLs and is therefore precisely quantized~\cite{Laughlin1981, Thouless1982, Hatsugai1993}. Such correlation between transport and the bulk topological order has been considered essential to the robustness of the transport quantization, i.e. insensitive to specific sample details. Based on such correlation between transport and bulk filling, transport measurements have been used to infer properties of the bulk. This approach, developed from study of 2DESs based on semiconductor heterostructures, has been widely adopted without much examination in QH studies of graphene~\cite{Novoselov2005, Zhang2005}. Although governed by the same basic QH physics, atomically thin systems such as graphene differ in important ways from semiconductor-based 2DESs. In particular, graphene's hard wall confining potential has been predicted to result in charge accumulation near physical edges under electrostatic gating~\cite{Silvestrov2008}, which has been used to explain observations in several transport~\cite{Vasko2010,Venugopal2011,Vera-Marun2013,Barraud2014} and imaging~\cite{Chae2012} experiments. In the QH regime where the edges play a crucial role as the connection between bulk state and transport~\cite{Halperin1982,Buttiker1988,Thouless1993,Tsemekhman1997,Weis2011}, how such charge accumulation affects transport quantization remains elusive~\cite{Vera-Marun2013,Barraud2014,Hettmansperger2012}. A comprehensive understanding requires a detailed study by both transport and bulk-sensitive techniques on the same gated hall bar. To this end, we combine a scanning probe technique, Microwave Impedance Microscopy (MIM), that probes local capacitance and conductivity~\cite{Kundhikanjana2011,Lai2011,Ma2015}, with simultaneous transport measurement, to directly examine the correlation between transport and bulk filling in graphene devices. We uncover a correlation that is very different from the commonly assumed picture: each transport plateau occurs at $\sim 90\%$ filling of the bulk LLs, when the bulk is still conductive. This points to a revised spatial configuration for QH transport quantization in gated films, which will impact many commonly used practices in QH transport analysis.

\begin{figure}
\includegraphics[width=1.0\columnwidth]{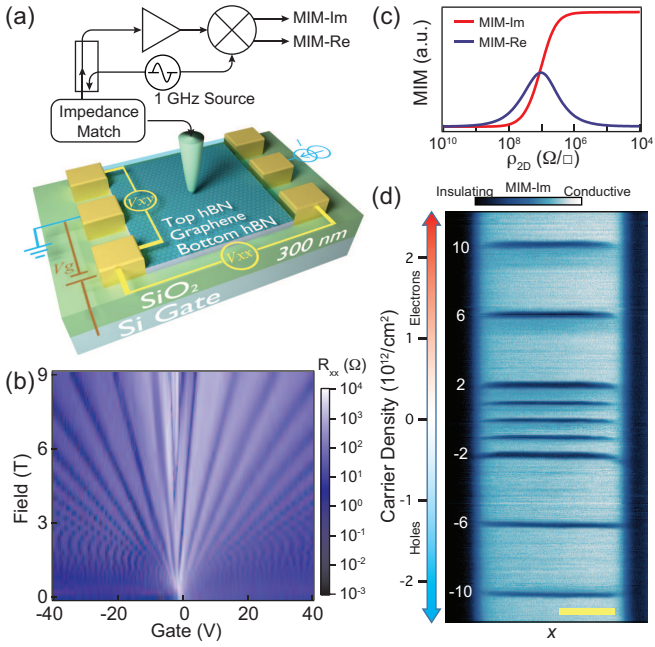}
\caption{\label{fig1}(color online) (a) Device structure and circuit schematic of MIM. (b) Landau fan diagram of device MLG \#1 at $T$ = 4.7 K. (c) Typical response curves of MIM-Im and MIM-Re as a function of 2D resistivity in a 2DES structure. The MIM-Im signal increases monotonically as the 2DES becomes more conductive and thus provides better screening, and the MIM-Re signal peaks at an intermediate resistivity value which maximizes power dissipation. (d) MIM-Im signals of repeated scans along the same line across MLG \#1 as the carrier density is tuned between p and n types at $B=$ 9 T and $T$ = 4.7 K. The suppression of the MIM-Im signal in the bulk corresponds to transitions through various Landau levels, matching the established Landau level structure in MLG. Scale bar is 2 $\mu$m.}
\end{figure}

We study three high-quality graphene devices: two monolayer graphene (MLG) and one bilayer graphene (BLG). In all cases graphene encapsulated above and below by hexagonal boron nitride (hBN) rests atop a dielectric layer of 300 nm SiO$_2$ on a conductive silicon substrate used as a back gate [Fig. \ref{fig1}(a)]. Electrical contacts to graphene are made via the recently developed one-dimensional edge metallization method~\cite{Wang2013,supp}. We present data from device MLG \#1 in the main text, and data from other devices exhibit qualitatively similar behavior and are provided in the Supplemental Material~\cite{supp}. These devices show well-developed quantized plateaux in transport at magnetic fields above 2 T at 5 K [Fig. \ref{fig1}(b)]. To probe the filling of LLs inside the graphene bulk, we perform MIM: spatially mapping tip-sample admittance (the inverse of impedance) as we raster a sharp metal tip over a graphene device. The schematic of the MIM measurement is shown in Fig. \ref{fig1}(a): a small microwave excitation (0.01-0.1 $\mu$W) at frequency of 1 GHz is delivered to a chemically etched tungsten tip. The reflected signal is amplified and demodulated into two output channels, MIM-Im and MIM-Re, which are proportional to the imaginary and real parts of the tip-sample admittance, respectively. Specifically, the imaginary component, MIM-Im, is a good measure of the local resistivity of the graphene as it decreases monotonically with increasing 2D resistivity [Fig. \ref{fig1}(c)]. When all LLs are filled or empty in the bulk, the local bulk resistivity is high, indicated by a low signal in the MIM-Im channel. Figure \ref{fig1}(d) shows the MIM-Im signal recorded during repeated scanning along a single line across device MLG \#1 as the gate is tuned from -40 V to 40 V. Indeed, a series of high-resistivity features matches the LL structure in MLG, i.e. at filling factor $\nu=\pm 2, \pm 6, \pm 10$, as well as the degeneracy-broken levels at $\pm 1$ and 0. 

\begin{figure}
\includegraphics[width=1.0\columnwidth]{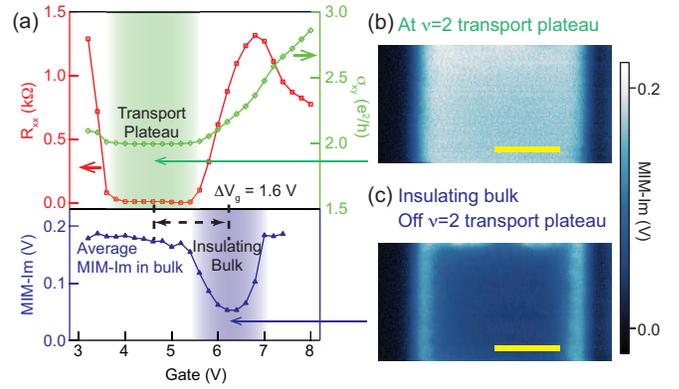}
\caption{\label{fig2}(color online) (a) Comparison of longitudinal resistance (red curve in the upper panel), normalized transverse Hall conductance (green curve in the upper panel), and MIM-Im signal averaged over a small region inside the graphene bulk (blue curve in the lower panel), measured at $T$ = 4.7 K and $B$ = 9 T for device MLG \#1. The shaded regions mark the transport plateau (green) and the incompressible bulk transition (blue), respectively. The difference in gate voltage between the center of the $R_{xx}$ minimum and the center of the MIM-Im minimum is 1.6 V. (b, c) Real space images of MIM-Im signal at gate values corresponding to the center of the transport plateau and the minimum of the bulk MIM-Im signal, as indicated by the arrows. Scale bar is 2 $\mu$m.}
\end{figure}

Next we examine the correlation between MIM signals in the bulk and simultaneously measured transport. Figure \ref{fig2}(a) shows an example of such correlation as the carrier density is tuned through the $\nu=2$ LL in MLG \#1. Surprisingly, we find that the range of gate voltages yielding a transport plateau has very little overlap with the range where the bulk is insulating. Such deviation is further revealed in the real space MIM-Im images: at the transport plateau where an insulating bulk surrounded by conductive edges is expected from the commonly assumed bulk/transport correlation, the MIM-Im image shows high conductivity in the graphene bulk [Fig. \ref{fig2}(b)]; the insulating bulk configuration only emerges at a higher gate value at which the transport is already exiting the plateau [Fig. \ref{fig2}(c)]. 

\begin{figure*}
\includegraphics[width=1.8\columnwidth]{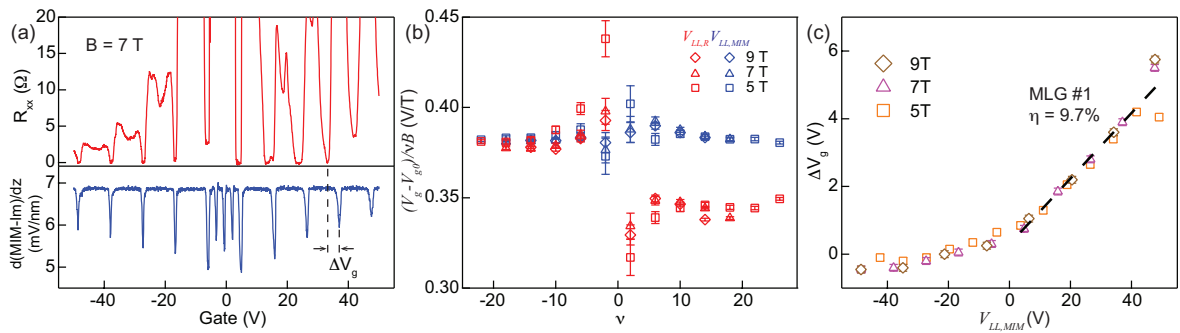}
\caption{\label{fig3}(color online) (a) Simultaneous measurement of $R_{xx}$ (upper panel) and  $d(\text{MIM-Im})/dz$ (lower panel) for MLG \#1 at $B$ = 7 T and $T$ = 4.7 K. (b) $(V_g-V_{g0})/\nu B$ plotted as a function of Landau level filling factor $\nu$. (c) $\Delta V_g$ plotted as a function of $V_{LL,MIM}$ which is proportional to the bulk density.}
\end{figure*}

Such deviation between transport plateau and insulating bulk occurs systematically as a function of carrier density and magnetic field. We specifically compare the MIM signal measured inside the bulk with $R_{xx}$ simultaneously measured through 4-terminal transport. In Fig. \ref{fig3}(a) we plot an example comparison of $R_{xx}$ and $d(\text{MIM-Im})/dz$ for MLG \#1. [$d(\text{MIM-Im})/dz$ behaves similarly to MIM-Im. See \cite{supp} for description of $d(\text{MIM-Im})/dz$ measurement and more data.] In both curves, a series of minima appear in $R_{xx}$ (matching plateaus in $R_{xy}$) and $d(\text{MIM-Im})/dz$. The corresponding gate voltages are defined as $V_{LL,R}$ and $V_{LL,MIM}$, respectively. To demonstrate that $V_{LL,MIM}$ is a consistent measure of the bulk LL filling, we analyze the normalized capacitance determined from these gate voltages. The bulk carrier density induced by a gate voltage, $V_g$, is $\rho_{bulk}=C_g (V_g-V_{g0})$ where $C_g$ is the graphene-to-gate capacitance per unit area and $V_{g0}$ the charge neutrality point. (The tip could also perturb the local carrier density due to a potential difference between tip and graphene. In our experiment, we have carefully checked the effect of tip bias and minimized it by applying a voltage on the tip to compensate the potential difference.) When these carriers completely fill LLs up to an integer filling factor  in the bulk, the density should be equal to the total density provided by the filled LLs, i.e. $C_g(V_g-V_{g0})=\nu\rho_0=\nu eB/h$ where $\rho_0$ is the density accommodated by a non-degenerate LL determined by the electron charge $e$, the magnetic field $B$, and the Planck constant $h$. This leads to a normalized quantity, $(V_g-V_{g0})/\nu B=e/hC_g$, which is expected to be a constant for a specific device~\cite{quantum_cap_note}. Figure \ref{fig3}(b) plots this quantity calculated from both $V_{LL, R}$ and $V_{LL,MIM}$ at different LLs. $V_{LL,MIM}$ provides consistent values across all LLs while the values calculated from $V_{LL,R}$ have large variations. This confirms that $V_{LL,MIM}$ corresponds to the complete filling of the bulk LLs. We now examine the deviation in gate voltage between a transport plateau and the complete filling of the corresponding bulk LL, i.e. $\Delta V_g=V_{LL, MIM}-V_{LL,R}$. $\Delta V_g$ is not zero for most LLs, and the deviations as a function of bulk state filling have approximately the same linear behavior independent of magnetic field, except that the electron and hole sides have different slopes. We believe that such electron-hole asymmetry is an extrinsic effect likely due to charge impurities near edges; thus the carrier polarity that has a larger slope would represent the intrinsic behavior in these devices. In fact, in one device with naturally occurring graphene edges fully encapsulated between hBN layers, such asymmetry is much reduced. (See discussions in ~\cite{supp}.) We find that the linearity slope $\eta$ for the $\Delta V_g$ versus $V_{LL,MIM}$ plot varies only slightly across the three devices, ranging between 7.8\% and 9.7\%~\cite{supp}. This value suggests that the transport plateau occurs at a bulk density $\rho_{bulk}=(1-\eta)\cdot \nu\rho_0 \approx 90\%\cdot \nu\rho_0$ for the LL at filling factor $\nu$.

To explain our experimental results, it is necessary to invoke a microscopic picture of QH transport quantization. There are currently two major models under active debate in the QH community: 1) the edge state picture in which dissipationless transport current is carried only by the edge states; 2) the incompressible state picture in which dissipationless transport current flows only in the incompressible regions where LLs are completely filled. Our results can be explained self-consistently within either model, and our conclusions on how to refine transport-based analysis of QH effect in graphene (which will be discussed later in the paper) do not depend on the model choice. To facilitate our discussion, we work in the incompressible state picture (for a review, see \cite{Weis2011} and references therein). The essence of this picture is that the dissipationless transport current flows only in the incompressible regions, hence the transport quantization depends on the spatial structure of such regions. In the incompressible regions, carriers cannot rearrange to screen electrostatic potential variations~\cite{Chklovskii1992,Yacoby1999}, allowing a transverse Hall potential drop~\cite{McCormick1999,Wei1998,Weitz2000}. The transverse electric field associated with this potential drop induces a drift velocity along the longitudinal direction for the occupied states, translating into a dissipationless current density. Integration of the current density over the entire incompressible region gives rise to a quantized transverse Hall conductance determined by the Landau level filling factor $\nu$ in the incompressible region. In a Hall bar geometry, the conditions to measure a quantized conductance $\nu h/e^2$ are: 1) the incompressible regions at filling factor $\nu$ are able to percolate through the sample and connect to the current leads, and 2) the voltage probes are able to achieve a full equilibration in chemical potential with the boundaries of these incompressible regions~\cite{Tsemekhman1997,Weis2011}.

\begin{figure}
\includegraphics[width=0.8\columnwidth]{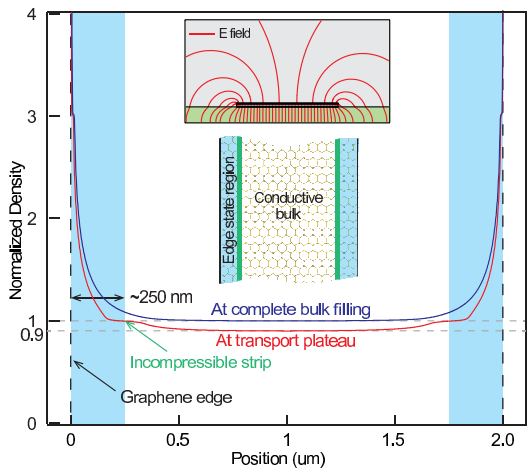}
\caption{\label{fig4}(color online) Simulated carrier density profiles corresponding to the transport plateau (red) of $\nu=2$, which is at 90\% filling of the bulk LLs, and the completely filled bulk (blue). The density is normalized by the density value corresponding to the complete filling of the bulk LLs, in this case, $2\rho_0$. The upper inset shows the electric field distribution. An incompressible strip forms near the position where the density crosses the complete filling value. The distance of the incompressible strip from the graphene edge is $\sim 250$ nm, independent of the particular LL. The lower inset shows the real-space conductivity configuration corresponding to the transport plateau: the conductive edge region is separated from the conductive bulk by the incompressible strip.}
\end{figure}

To explain the observed deviation between transport and bulk filling, it is important to examine the spatial variation of carrier density, especially near the edges. Due to the hard wall confining potential of graphene, electrostatic back gating will naturally induce more charges near edges than in the bulk~\cite{Silvestrov2008}. We simulate this effect using finite element software and take into account the LL energy structure of MLG. Example carrier density profiles are plotted in Fig. \ref{fig4} for the case of $\nu=2$. For the profile corresponding to the transport plateau where $\rho_{bulk}=90\%\cdot 2\rho_0$ (the red curve), the carrier density is at 90\% filling in the bulk and increases to cross $2\rho_0$ at $\sim 250$ nm away from the physical edge. Around the crossing position, an incompressible region forms where the density stays constant at $2\rho_0$~\cite{Chklovskii1992,Weis2011,Yacoby1999}. Since transport quantization is achieved in our experiments, transverse Hall voltage drops must be established across the two incompressible strips at $2\rho_0$, one near each edge, and under this same gating condition the voltage probes must be able to establish a full equilibration with the boundaries of the incompressible strips. This scenario is similar to the higher-than-integer bulk filling (lower magnetic field) side of a QH plateau in a GaAs-based 2DES~\cite{Weis2011}. In this graphene system, upon further filling the incompressible strips will move into the bulk, and eventually merge when the bulk reaches complete filling. The charge accumulation region will become considerably wider (see the blue curve in Fig. \ref{fig4}). We suspect that the reason for losing transport quantization is because additional incompressible strips can emerge at positions where the accumulated carrier density reaches the next LL near the edge, thus perturbing the equilibration between the voltage probes and the $\nu=2$ incompressible bulk.

Since the electrostatically induced density profile scales linearly with the back gate voltage, the profile for the $\nu$-th transport plateau, when normalized by the corresponding value of the bulk density, $\nu\rho_0$, has the same shape as that for $\nu=2$: the bulk density is $90\%\cdot\nu\rho_0$ and the profile crosses the complete filling of $\nu\rho_0$ around the same position $\sim 250$ nm away from the physical edge. The only difference from one transport plateau to the next is that the width of the incompressible strip also depends on the energy gap which varies at different LLs. Based on the electrostatic modeling and previous analysis, the observation of $\eta=90\%$ implies that the condition to achieve the QH transport quantization at filling factor $\nu$ in gated graphene devices is when the gate-induced carrier profile crosses integer filling $(\nu\rho_0) \sim 250$ nm away from the physical edge so that only one incompressible strip forms near each edge. The next higher density incompressible strip would be so close to the edge that the potential slope would be very steep, so the strip would be very narrow and easy to tunnel across –-- the incipient incompressible strip at $6\rho_0$ (normalized density 3) is barely visible in the simulation in Fig. \ref{fig4}.

In the following we discuss the implications of the observed correlation between transport and bulk filling on a variety of experiments in the study of the QH effect in graphene as well as transport in other related 2D topological systems. The spatial configuration of incompressible regions at QH transport quantization is the basis for analyzing many transport results. The well-established correlation between transport plateau and bulk filling based on studies in semiconductor-based 2DESs has been widely adopted in the QH community and naturally assumed to apply for graphene devices. However, our experiment reveals that the picture in gated graphene devices with a dielectric layer of 300 nm SiO$_2$, the most widely used geometry in graphene research, deviates from the conventional correlation in a non-trivial way: the QH transport quantization occurs when the bulk is still conductive with partially filled LLs. Therefore, many common practices used in the QH study on such device geometry need to be re-examined for validity. 1) For example, the conventional method of estimating the bulk carrier density based on QH plateau positions likely needs to be corrected. 2) Transport-based analysis of dissipation could be affected. Transport dissipation can occur through tunneling across the incompressible region. In the commonly assumed spatial configuration, a fully incompressible bulk is considered around the center of the QH plateau, so the tunneling probability is expected to be exponentially small and is often neglected. Our results suggest that the incompressible region in gated graphene devices is considerably narrower at the QH plateau so that tunneling across such regions must be included in the analyses. For example, extraction of bulk LL gaps through thermal activation analysis may not be accurate, because this method assumes that carriers in the incompressible region are thermally excited to fill the available states in the next LL, neglecting tunneling across the broad incompressible region~\cite{Giesbers2007}. Our results can also help guide which regime to use for metrology applications where QH transport in graphene devices is used as a resistance standard~\cite{Giesbers2008,Tzalenchuk2010,Ribeiro-Palau2015}. Achieving high accuracy of QH resistance quantization requires minimization of dissipation, favoring wide incompressible regions. Our results suggest that the edge carrier profile should be carefully considered when gated graphene devices are used for such purpose~\cite{Giesbers2008} because it can alter the configuration of incompressible regions at QH plateaux. 3) Transport-based analysis of localization during the transition between LL plateaux may need to be revisited since the evolution of the bulk state during the transition is different from the conventional assumption. For example, the validity of the variable range hopping model~\cite{Bennaceur2012} needs to be re-examined because the bulk localization occurs at a density value away from the transport plateau.

Understanding edge behaviors also has implications for transport studies on other 2D topological effects, such as the quantum spin Hall~\cite{Bernevig2006} and quantum anomalous Hall~\cite{Yu2010} systems. While signatures of edge transport have been identified in experiments~\cite{Konig2007,Du2015,Chang2013}, it is important to distinguish between non-trivial edge states that have a topological origin and trivial edge states that could also contribute to transport and even generate transport signatures similar to those expected for topological modes~\cite{Ma2015,Nichele2015}. The electrostatically-induced carrier profile that peaks near the edge is not unique to graphene, so its effect on topological transport in other similar systems needs to be taken into account and carefully analyzed.
Such an edge effect may also enable new device applications. Our result suggests ways to manipulate the edge density profile to control the QH transport quantization, such as using multiple side/back gates to control electrostatic potential profile at the graphene edge~\cite{Panchal2014}. A recent work proposes the idea of guiding electron waves based on an edge potential in graphene~\cite{Allen2015}. Electrostatic gating thus offers a convenient way to control and manipulate the edge potential.

To summarize, through a comprehensive study combining transport and MIM measurements, we observe that QH transport plateaux occur at $\sim$90\% filling of the bulk LLs in gated graphene devices. Such unconventional correlation between transport quantization and bulk state filling provides a basis to understand the QH transport quantization in graphene, and also has important implications on transport studies in other 2D topological systems.

We thank Xudong Chen for help with the COMSOL simulation. The work at Stanford University is supported by the Gordon and Betty Moore Foundation through Grant No. GBMF3133 and the Emergent Phenomena in Quantum Systems (EPiQS) initiative GBMF4546 to Z.-X.S., and through Grant No. GBMF3429 to D.G.-G. This work is also supported by the NSF Grant DMR-1305731. The work at Columbia University is supported by the NSF Grant DMR-1463465.

%\bibliography{Graphene_QH_carrier_profile}

\begin{thebibliography}{42}
\bibitem [{qua()}]{quantum_cap_note}%
  \BibitemOpen
  \href@noop {} {}\bibinfo {note} {We ignored contribution from quantum
  capacitance because it is expected to be much larger than geometrical
  capacitance for devices with a $\sim 300$ nm SiO$_2$ dielectric.}\BibitemShut
  {Stop}%
\end{thebibliography}

\end{document}